\def\be{\begin{equation}}
\def\ee{\end{equation}}
\def\bea{\begin{eqnarray}}
\def\eea{\end{eqnarray}}
\def\mk{{\mathbf k}}

\def\bra#1{\langle #1 |}
\def\ket#1{|#1 \rangle}
\def\braket#1#2{\langle #1 | #2 \rangle}
\def\Pa{{\cal P}(\alpha)}
\def\Da{D(\alpha)}
\def\Sz{S(z)}
\def\ave#1{\langle #1 \rangle}

\documentclass[twocolumn,twoside,preprintnumbers,amsmath,amssymb,showkeys,showpacs,floatfix]{revtex4}
\usepackage{epsfig,graphicx,fancyhdr}

\begin{document}
\title {Disappearance of Squeezed Back-to-Back Correlations -
        a new signal of hadron freeze-out from a supercooled Quark Gluon Plasma}

\author{T. Cs\"org\H{o}$^{1,2}$ and Sandra S. Padula$^1$ }

\affiliation{
        $^1$Instituto de F\'\i sica Te\'orica - UNESP, \\
        Rua Pamplona 145, 01405-900 S\~ao Paulo, SP, Brazil, \\
         $^2$ MTA KFKI RMKI, H - 1525 Budapest 114, P.O.Box 49, Hungary
    }

\begin{abstract}
We briefly discuss four different possible types of transitions
from quark to hadronic matter and their characteristic signatures
in terms of correlations. We also highlight the effects arising
from mass modification of hadrons in hot and dense hadronic
matter, as well as their quantum statistical consequences: the
appearance of squeezed quantum states and the associated
experimental signatures, i.e., the back-to-back correlations of
particle - anti-particle pairs. We briefly review the theoretical
results of these squeezed quanta, generated by in-medium modified
masses, starting from the first indication of the existence of
surprising particle - anti-particle correlations, and ending by
considering the effects of chiral dynamics on these correlation
patterns. A prerequisite for such a signature is the experimental
verification that these theoretically predicted back-to-back
correlation of particle anti-particle pairs are, in fact,
observable in high energy heavy ion reactions. Therefore, the
experimental observation of back-to-back correlations in high
energy heavy ion reactions would be a unique signature, proving
the existence of in-medium mass modification of hadronic states.
On the other hand, their disappearance at some threshold
centrality or collision energy would indicate that the hadron
formation mechanism would have qualitatively changed: asymptotic
hadrons above such a threshold are not formed from medium modified
hadrons anymore, but rather by new degrees of freedom
characterizing the medium. Furthermore, the disappearance of the
squeezed BBC could also serve as a signature of a sudden,
non-equilibrium hadronization scenario from a supercooled
quark-gluon plasma phase. 
\end{abstract}

\vskip -1.35cm

\maketitle

\thispagestyle{fancy}

\setcounter{page}{1}
\bigskip

\bigskip

\section{INTRODUCTION}
\label{s:intro} The major goal of the programme of high energy
heavy ion physics has been to explore the phases of hot and dense,
strongly interacting matter, more specifically, to find a new
phase where quarks and gluons are no longer confined to their $T=
0$ bound states, the hadrons. This new phase of matter may be
referred to as ``Quark-Gluon Plasma'' (QGP) in analogy to the
electromagnetic plasma, the phase in which electrons and ionized
atoms appear as new degrees of freedom, as compared to the neutral
atomic or molecular forms of matter.

It has been theoretically predicted by a number of
authors~\cite{Pratt:1984su,Pratt:1986cc,Hama:1987xv,Bertsch:1988db,Bertsch:1989vn,Rischke:1996em}
that the emergence of a large number of colored degrees of freedom
in an expanding QGP could be signaled by correlation measurements.
However, some of the theoretical expectations related to a first
order phase transition, such as a long-lived QGP, are not being
observed experimentally.

Nevertheless, we could still ask the question: can a phase
transition be signaled by correlation measurements {\it at all}?
Fortunately, the answer is positive: In section ~\ref{s:evidence},
we shall discuss correlation measurements that have indeed
signaled experimentally the onset of first order as well as second
order phase transitions, in solid-state physics measurements.

\section{PHASE TRANSITIONS AND CORRELATIONS}
\label{s:evidence}

Can an experimental measurement of the two-particle correlation
function be utilized to signal a phase transition? A clear yes 
answer has recently been given in condensed matter
physics by Schellekens, Hoppeler, Perrin, Viana Gomes, Boiron,
Aspect and Westbrook~\cite{Schellekens2005}. In their measurement,
a cloud of ultra-cold gas of meta-stable He atoms was released
from its magnetic trap, and after 47 cm ballistic fall, the
position of each atom was detected. Single particle detection of
these neutral atoms was possible as the 20 eV internal energy of
each atom was released when the atoms hit the detector.

The initial temperature has been varied from  above to below the
Bose-Einstein condensation threshold, $T_{BEC} = 0.5$ $ \mu K$.
The observed two-body correlation function for initial conditions
with $T > T_{BEC}$ showed a bunching behavior, while the
correlation function became flat for the coherent, Bose-Einstein
condensed sample, with initial conditions $T < T_{BEC}$. The
observed quantum statistical correlations are the atomic analogue
of the Hanbury Brown -- Twiss effect~\cite{HBT}, and in the case
of atomic Bose-Einstein condensation, the disappearance of the
bunching behavior signaled the phase transition from the usual
thermal state to the Bose-Einstein condensate. The intercept
parameter of the correlation function served as an order parameter
of this phase transition. 
Thus the 
correlation function can signal a phase transition clearly.

\section{FOUR POSSIBILITIES FOR TRANSITION FROM DECONFINED TO HADRONIC MATTER}
\label{s:fourtr}

What kind of phase transitions may occur in hot and dense, nuclear matter?
According to recent lattice QCD calculations,
the three kind of equilibrium transitions are possible on the ($T$, $\mu_B$) plane:
 if the temperature is increased
at zero, or nearly zero chemical potentials, the transition from confined
to deconfined matter is a cross-over~\cite{Fodor-Katz-crossover},
 i.e. this is not  a phase transition
in the strict, thermodynamical sense: quark and gluon degrees of
freedom are present below $T_c$, hadrons are present above $T_c$
and various observables yield different estimates for the critical
temperature itself. In contrast, at rather large bariochemical
potentials, the transition from hadronic matter to deconfined
matter is of first order. The line of first order transitions ends
at a critical end point (CEP) that separates the first order phase
transitions from the cross-over region. At this critical end
point, $(\mu_{CEP}, T_{CEP})$ , the transition from confined to
deconfined matter is a second order phase transition. There are
other more exotic states like color superconducting quark matter,
or there is a nuclear liquid-gas phase transition, but the region
of the phase diagram that is relevant for high energy heavy ion
collisions at CERN SPS, RHIC and LHC, is the region around the
CEP~\cite{Fodor-Katz-CEP}. Recent calculations and experimental
data indicate that in central Au+Au collisions at RHIC, the
transition is likely a cross-over:
The
extracted central temperatures for the 0-5 \% most central
$\sqrt{s_{NN}} = 130$ GeV Au+Au data sample are $T_0 = 214 \pm 7 $
MeV, and $\mu_B = 77 \pm 38$ MeV, respectively~\cite{BL-indication}.
These values are to be contrasted also with the location of the
critical end point of the line of the 1st order QCD phase
transition, located preliminarily at $T_{CEP} = 162 \pm 2$ MeV and
$\mu_{CEP} = 360 \pm 40$ MeV, according to the lattice QCD
estimate~\cite{Fodor-Katz-CEP}.

If the evolution of the strongly interacting matter produced in
relativistic heavy ion collisions happens through
thermodynamically equilibrated states, first- and second-order
phase transitions, as well as an analytic cross-over, are the only
three possibilities for the state of the system. However, heavy
ion collisions create violently exploding mini-bangs, where the
time-scales of the expansion are rather short as compared to the
typically $\sim 100$ fm/$c$ nucleation times of hadronic bubbles
in a first order phase transition from  QGP to a hadron gas. Hence
non-equilibrium transitions are also possible. In
ref.~\cite{Csorgo:1994dd}, another scenario is suggested, where a
rapidly expanding QGP state might strongly supercool in Au+Au
collisions at RHIC and then hadronize suddenly while emitting a
flash of pions. The predicted signals of this scenario are not,
even currently, in disagreement with RHIC data on Au+Au
collisions. So the possibility of non-equilibrium hadron formation
must be kept in mind and further experimental tools must be found
to pin down if this mechanism is in fact present in Au+Au
collisions at RHIC.

\section{CORRELATION SIGNATURES FOR THE FOUR REHADRONIZATION SCENARIOS}
\label{s:four-corr}
Let us itemize the possible types of rehadronization transitions,
and summarize their correlation signatures:

\begin{itemize}
\item A strong 1st order QCD phase transition
 has been studied in the predominant fraction of the literature. Such a
transition is characterized by long life-times and 
time spreads, since the latent heat and the initial entropy
densities are large. The picture behind this scenario is  that of
a slowly burning cylinder of QGP~\cite{Rischke:1996em}. In such a
scenario, the extended duration of particle emission adds  up to
source parameter in the direction of the average momentum of the
pion pair, resulting in a substantial increase of the effective
source size. This is signaled as $R_{out} \gg R_{side}$,
regardless of the exact details of the calculation
~\cite{Pratt:1984su,Pratt:1986cc,Hama:1987xv,Bertsch:1988db,Bertsch:1989vn,Rischke:1996em}.
However, experimentally $R_{out} \approx R_{side}$ both at CERN
SPS ~\cite{NA44,NA49} and at
RHIC~\cite{STARout,PHENIXout,PHOBOSout}.
 Hence, a strong first order transition from deconfined  to hadronic matter
seems to be excluded by current correlation measurements in high
energy heavy ion collisions, both at CERN/SPS and at the RHIC
energy range.
\end{itemize}

However, three other alternatives still remain.

\begin{itemize}

\item A second order QCD phase transition from deconfined to
hadronic matter has been considered recently in
ref.~\cite{csorgo-hegyi-novak-zajc}. The interesting conclusion is
that this phase transition is not signaled by the scale parameters
of the correlation function, as the spatial correlations develop a
power-law tail in these kind of transitions, and power-laws have
no characteristic scale. Instead, the second order phase
transitions are characterized by critical exponents. One of these,
traditionally denoted by $\eta$, characterizes the tail of the
spatial correlations of the order parameter. In
ref.~\cite{csorgo-hegyi-novak-zajc}, this exponent was shown to be
measurable with the help of the two-particle correlation function
in momentum space, and it was shown that $\eta = \alpha$, where
$\alpha$ stands for the L\'evy index of stability of the
correlation function itself, i.e., $C(q) = 1 + \lambda \exp[ - (q
R)^\alpha ]$.

A strong decrease of the L\'evy index of stability $\alpha$ down
to about $0.50 \pm 0.02$ in the vicinity of the critical end point
is a 
theoretical prediction for the localization of this outstanding
landmark of the QCD phase diagram. This value was obtained based
on universality class arguments. Rajagopal, Wilczek and others
argued that the universality class of QCD at the critical point is
that of the 3-dimensional Ising model ~\cite{Rajagopal-Wilczek}.
The exponent of the correlation function $\eta $, however, is
extremely small in that model, i.e., $\eta = 0.03 \pm
0.01$~\cite{Rajagopal-Wilczek}. In a violent heavy ion collisions,
random external fields are also present, which change the
universality class and, thus, might increase the value of the
correlation exponent. In the 3-dimensional random field Ising
model, $\eta$ increases to $0.50 \pm 0.05$. When interpreted as a
L\'evy index of stability, it still corresponds to an extremely
peaked correlation function, which should be very clearly
measurable. Nevertheless, the excitation function of the {\it
shape parameter} $\alpha$ has
not yet been determined experimentally. 
Therefore, it is possible that at certain colliding energies,
either at CERN/SPS range or at RHIC, there is a dramatic change in
the shape of the correlation function, which has not yet been
identified. 

\item A cross-over 
from QGP to confined matter is the last remaining possibility of
an equilibrium phase transition. This scenario can be signaled by
emission of hadrons from a region above the critical temperature,
$T > T_c$, or, by finding deconfined quarks or gluons at
temperatures $T < T_c$. Recent lattice QCD calculations suggest
that, at zero baryochemical potential, the transition from quarks
to hadrons is a rapid cross-over, different observables yielding
different transition temperatures at $\mu_B= 0$. The peak of the
renormalized chiral susceptibility predicts $T_c$=151(3) MeV,
whereas critical temperatures based on the strange quark number
susceptibility, and Polyakov loops, result in values higher than
this by $24(4)$ MeV and $25(4)$ MeV, respectively. Signs of quarks
or gluons below these temperatures have not yet been observed
experimentally. However, emission of hadrons from a small but very
hot region, with $T > T_c \simeq 176 \pm 7$
MeV~\cite{Aoki:2006br,Fodor-Katz-CEP}, has been suggested by
Buda-Lund hydro model fits to the identified particle spectra and
two-pion correlation functions in Au+Au collisions at
RHIC~\cite{BL-indication}.

\item
The fourth possibility corresponds to hadron formation out of
thermal equilibrium. A sudden recombination from quarks to hadrons
has been considered as the mechanism for hadron formation in the
ALCOR model~\cite{ALCOR}. Other realizations of sudden hadron
formation and freeze-out were tools used to explain the observed
scaling properties of elliptic flow~\cite{v2-scaling} with the
number of constituent quarks. In the context of femtoscopy
analyses, it has been suggested that the comparable magnitude of
$R_{out} \approx R_{side} \approx R_{long}$ could be a signature
of a flash of hadrons (pion-flash) from a deeply supercooled QGP
phase~\cite{Csorgo:1994dd}. If this mechanism would indeed be
responsible for the hadronization in high energy heavy ion
collisions, it would also predict that the freeze-out
distributions of temperature, flow, and density would not depend
on the particle type (or cross-section), since all hadrons would
be produced in the same flash. Strangeness would be enhanced, as
strangeness production predominantly happens in the pre-hadron
state of matter. Finally, no in-medium mass modification of
hadrons could be observed, since deeply supercooled QGP is a state
of over-stretched matter, has negative pressures internally, which
cause the sudden break-up and coalescence of matter. However, in
this picture, the hadron gas is produced in a large volume, and
re-scattering effects are small. Hence, the interactions which
would cause the in-medium mass modification of hadrons, do not
take place, as the hadron gas would freeze out as soon as it was
produced. Note, however, that the thermodynamic considerations in
Ref. ~\cite{Csorgo:1994dd} would also allow, in about 50 \% of the
parameter space, for the production of a super-heated hadron gas.
Therefore, the indication of a hot spot in the central region, and
of particle emission from a region with $T(x) >
T_c$~\cite{BL-indication} would be compatible with this
non-equilibrium hadronization mechanism as well.

\end{itemize}

In the end of this review, we shall propose a new method to
observe the onset of sudden hadronization from a supercooled QGP
state as a possible rehadronization mechanism.

\section{COHERENT STATES\label{s:coh}}

In this session, we will briefly present the theory of quantum
optical coherence and chaos, by introducing the concept of
coherent states and an explanation of the HBT effect. Then we
discuss squeezed states and review their applications in high
energy heavy ion and particle physics. At the end we will focus on
how these quantum statistical correlations of squeezed hadronic
states could be experimentally used as tools to search for a
sudden freeze-out of hadrons from a super-cooled QGP state.

 Let us mention an inspiring historical review by Michel Martin
Nieto~\cite{Nieto:1997xf}, who compared the well known discovery
of the coherent states by E. Schr\"odinger in
1926~\cite{Schrodinger1926} with the much less well known
discovery of squeezed states by E. H. Kennard in
1927~\cite{Kennard1927}. According to Nieto~\cite{Nieto:1997xf},
Schr\"odinger's original discovery of coherent states was inspired
by a question from Lorentz, in a letter on May 27, 1926, in which
he lamented the fact that Schr\"odinger's wave functions were
stationary, and did not display classical motion. On June 6, 1926
Schr\"odinger replied that he had found a system with classical
motion, and sent Lorentz a draft copy of his paper, published in
Ref.~\cite{Schrodinger1926}. According to it, coherent states can
be defined with the help of the so-called minimum-uncertainty
method. In this, the mean position and mean momentum are required
to follow the trajectory of the classical motion corresponding to
the same Hamiltonian, and also that the vacuum state is a member
of the set of states.

The development of understanding about the dual nature of light,
evident in its wave-like properties and quantized detections has
been summarized recently in ref.~\cite{Nobel2005}. 

When the tools to handle quantum electrodynamics were developed,
they were applied mainly to high energy processes, and it was
still naively assumed, that the conflict between Maxwell's
treatment - who focused on the wave-like properties of the
electromagnetic fields, and whose formulas were the basis of radio
engineering - and Planck's theory, which emphasized the quantum
nature of light, would be of no significance in optical
observations. This state of blissful indifference~\cite{Nobel2005}
was however changed by the landmark experiment of R. Hanbury
Brown and R. Q. Twiss~\cite{HBT}, who 
proposed an interferometric method to determine the angular
extension of distant stellar objects. 
They found out that intensity correlation between photocurrents,
recorded in two separated detectors, displayed a bump when the
difference in optical path lengths between the signals was
decreased towards zero. These observations have shown that the
photons in the two different detectors were correlated, although
they stemmed from two different surface elements of distant stars.
This way, individual photons, as well as pairs of such photons,
entered the realm of observational optics. Ever since, this
phenomenon is known as the HBT effect.

The correlated emission of particle pairs is a fundamental
property of quantum fields. The effect has been observed in
particle physics and been explained based on the bosonic nature of
pions  by Goldhaber, Goldhaber, Lee, and Pais in 1960~\cite{GGLP}.
Recently, such positive correlations have been observed also in
ultracold  quantum gases by Schellekens et
al~\cite{Schellekens2005}, as detailed in Section~\ref{s:evidence}, which indicates
that quantum fluctuations and correlations are important basic
properties of matter fields, just as well as that of the
electromagnetic (photon) field~\cite{Knight2005}. 

After the discovery of the HBT effect and the discovery of lasers,
there was a theoretical debate in the literature, arguing if
photons from a laser light would show the HBT effect or not. The
correct theory was published first by Glauber~
who had shown that there is no HBT effect in the coherent fields
of lasers. In fact, the lack of second- and higher-order HBT type
of intensity correlations {\it defines} quantum optical coherence.
Thus, Glauber related the bunching properties of the HBT
observations to the  random, chaotic nature of the photon field in
thermal radiation. He also pointed out that more information is
necessary to characterize the quantum state of the photon field:
{\sl ``Whereas a stationary Gaussian stochastic process is
described completely by its frequency-dependent power spectrum, a
great deal more information in the form of amplitude and phase
relations between differing quantum states my be required to
describe a steady light beam. Beams of identical spectral
distributions may exhibit altogether different photon correlations
or, alternatively, none at all. 
There is ultimately no
substitute for the quantum theory in describing quanta 
''}. Based on Glauber's theory, the puzzling HBT effects observed
in thermal radiation and the lack of HBT effect in lasers were
understood in the same framework, and the characterization of
quantized electromagnetic radiation has been reduced to counting
photons. The new field of {\it quantum optics} was born. These
days, we witness the birth of {\it quantum
atomics}~\cite{Knight2005} as coherent and incoherent beams of
fermionic ~\cite{Henny1999} or bosonic~\cite{Schellekens2005}
atoms enter the realm of experimental investigations.

In this section, we highlight some of the basic properties of
coherent states, introduced by Glauber in
Ref.~\cite{Glauber:1962tt} as eigenstates of the annihilation
operator. For simplicity, let us consider a given mode of a free
boson field, corresponding to a harmonic oscillator. After
quantization and normal ordering, the Hamiltonian operator of the
one-mode harmonic oscillator is written as \be
        H = \omega a^\dagger a,
\ee where $a^\dagger$ and $a$ are the creation and annihilation
operators, respectively. Their non-vanishing commutator is
 \be
        [a, a^\dagger] =  1.
\ee
With the help of these creation and annihilation operators, a
Fock space can be built up by considering that 
\bea
        a \ket{0} &  = & 0,\\
        \ket{n} & = & \frac{1}{\sqrt{n!}}\;a^{\dagger n} \ket{0}.
\eea
The creation and annihilation operators step up or down on
the infinite ladder of Fock spaces as \bea
        a^\dagger \ket{n} & = & \sqrt{n+1}\;\ket{n+1}, \\
        a \ket{n} & = & \sqrt{n}\;\ket{n-1}.
\eea Here $n$ stands for a non-negative integer number. The number
operator is the Hermitian $N = a^\dagger a$, with $N \ket{n} = n
\ket{n}$. These Fock states are orthonormal, \be
        \braket{n}{m} = \delta_{n,m} ,
\ee where $\delta_{n,m} $ stands for a Kronecker-delta. The
resolution of the unity operator in terms of Fock states is given
by
 \be
        1 = \sum_{n = 0}^\infty \ket{n}\bra{n}.
\ee So, these Fock states form a complete, orthonormal basis. The
density matrix of the system can be expanded as \be
        \rho = \sum_{n= 0}^{\infty} p_n \ket{n}\bra{n},
\ee where $p_n$ stands for the probability that the quantum
mechanical system is in state $\ket{n}$. Its normalization is
 \be
        \mathrm{Tr} \rho = \sum_{n=0}^\infty p_n = 1 .
\ee
The above formulas can be generalized with the help of coherent states,
but with certain subtleties. Let us start first with definitions.

Coherent states are the eigenstates of the annihilation operator,
\be
        a \ket{\alpha} = \alpha \ket{\alpha} ,
\ee
where $\alpha $ is a complex (c) number.
Their algebraic properties are very interesting.
For example, one can express them in terms of Fock states as
\be
        \ket{\alpha } = \exp( - |\alpha|^2 /2 ) \sum_{n=0}^\infty
\frac{\alpha^n}{\sqrt{n!}} \, \ket{n} . \ee

It is interesting to observe that, removing (detecting) one
quantum from a coherent state does not change the probability that
yet another quantum could be removed from such a state. This is in
sharp contrast to the properties of Fock states where, once a
particle is detected, the resulting Fock state is orthogonal to
the Fock state before the detection. i.e., $\braket{n}{n-1} = 0$.

The expansion of $\ket{\alpha}$ in terms of Fock states can be
used to prove that these states are properly normalized,
$\braket{\alpha}{\alpha} = 1$. Although different coherent states
are not orthogonal, i.e.,
\be
        |\braket{\alpha}{\beta}|^2 = \exp( - |\alpha - \beta|^2),
\ee
they can, nevertheless, also be used to resolve unity as \be
        1 = \int \frac{{\mathrm d}^2\alpha}{\pi} \ket{\alpha}        \bra{\alpha},
\ee where $d^2\alpha = d {\mathrm Re}[\alpha] \, d {\mathrm Im}
[\alpha]$. It then follows that any state may be expanded linearly
in terms of coherent states. Therefore, the most general light
beam (of a given mode ${\bf k}$ and of a given polarization) can
be described by a density operator of the following form: \be
\hat\rho = \int d^2\alpha d^2 \alpha^\prime {\cal P}_2(\alpha,
\alpha^\prime)
        \ket{\alpha}\bra{\alpha^\prime}.
\ee

In many practical important cases, such as the thermal or the
coherent radiation, the density matrix turned out to be in a
diagonal representation when expanded in terms of the overcomplete
set of coherent states:
\be
    \hat\rho = \int d^2\alpha
        {\cal P}(\alpha)
        \ket{\alpha}\bra{\alpha}.
\ee
A similar representation was proposed by
Sudarshan~\cite{Sudarshan:1963ts} shortly after Glauber's work,
which emphasized the advantage of this representation when
evaluating expectation values for normal ordered products of
creation and annihilation operators,
\be {\rm Tr} \hat\rho \hat O
= {\rm Tr} \left\{\hat\rho  (a^{\dagger})^\lambda a^\mu \right\}=
        \int d^2\alpha {\cal P}(\alpha) (\alpha^*)^\lambda
        \alpha^\mu.
\ee

Consequently, the evaluation of expectation values of normal
ordered operators in a quantum mechanical system, defined with the
help of its density matrix, is reduced to the evaluation of
expectation values of the quasi-probability distribution $\Pa$,
defined over the complex plane of $\alpha$. The Hermiticity of the
density matrix implies that $ \Pa $ is a real valued function,
although it is not necessarily positive, due to the quantum nature
of the field. This representation of the density matrix is
frequently called as 
        the Glauber-Sudarshan representation.

Glauber investigated a model of photoionization of a pair of
atoms, labeled 1 and 2, lying at ${\bf r}_1$ and ${\bf r}_2$
within a light beam of sufficiently narrow spectral bandwidth and
given polarization. Summing up the transition probabilities over
final electron energies, there is no quantum mechanical
restriction in defining the time at which each electron emission
takes place. The probability density for ionization of atom 1 at
time $t_1$ and for atom 2 at time $t_2$ can be written as \be
        P_2(t_1,t_2) = P_1(t_1) P_1(t_2) C_2(t_1,t_2) ,
\ee
where $P_1(t)$ is the transition probability
of each atom placed  individually in the beam, and $C_2(t_1,t_2)$ is the
two-particle correlation function.

Glauber pointed out, that for a coherent state, corresponding to
$\Pa = \delta(\alpha - \beta) $, $C_2(t_1,t_2) = 1$, i.e., there
is no HBT type of correlation in a coherent beam, which would be a
reasonable model for the laser field. The density matrix of a
thermal radiation with mean number of photons $\ave{n}$ can be
written as
\bea
\hat \rho & = &\sum_{n=0}^\infty p_n \ket{n}\bra{n},\\
p_n & = & \frac{\ave{n}^n }{(1 + \ave{n})^{1 + n}}. \eea
This corresponds to a Gaussian distribution in terms of the $\Pa$
representation, \be \Pa = \frac{1}{\pi \ave{n}} \exp( -
|\alpha|^2/ \ave{n}). \ee It is easy to show that for such a
filtered, single mode, thermal radiation, the correlation function
is
\be
        C_2 = 1 + |\exp\left( i \omega [t_2 - t_1 -(x_2 - x_1)/c] \right) |^2.
\ee
This corresponds to the classical limit, when modes are
treated as forming a continuum with a stochastic noise.
Note that for such a thermal light, the maximum value of the correlator is
 $C_2 = 2$, in a sharp contrast to the $C_2(t_1,t_2) \equiv 1$ value,
that is characteristic  of a coherent light.

\section{SQUEEZED CORRELATIONS IN PARTICLE FEMTOSCOPY}
\label{s:squeezed-review}

Squeezed states are generalized coherent states with very
interesting quantum statistical properties. They can be obtained
by means of the displacement operator method, as discussed below,
by using a Bogoliubov-Valatin transformation, or even as
generalized minimum uncertainty states~\cite{Nieto:1997xf}. Let us
consider first the displacement operator method, and discuss their
properties in terms of Bogoliubov transformation in the next
subsection.

\subsection{SQUEEZED COHERENT STATES}
\label{ss:squeezed-intro}
\noindent The displacement operator has the following properties
\bea
        D^{-1}(\beta) a D(\beta) & = & a + \beta, \\
        D^{-1}(\beta) a^\dagger D(\beta) & = & a^\dagger + \beta^*.
\eea It is easy to show that acting with the first of the above
equality on a vacuum state, the displacement operator $\Da$
creates a coherent state, i.e.,
\be
        \Da \ket{0} = \ket{\alpha}
\ee On imposing the requirement that $\Da$ should be a unitary
operator, this leads to \bea
        \Da &=& \exp(\alpha a^\dagger - \alpha^* a) \nonumber \\
        &=&        \exp(-|\alpha|^2/2) \exp(\alpha a^\dagger) \exp(-\alpha^* a),
\eea where the second part expresses the displacement operator in
a normal order, 
i.e., where all the creation operators stand to the left 
and all the annihilation operators stand to the right. 
Further discussion of the properties of these displacement
operators, can be found in
Refs.~\cite{Glauber:1963fi,Glauber:1963tx}.

The displacement operator is generalized to the squeeze operator $S(z)$ as 
\bea
        S(z) & = & \exp\{\frac{1}{2}( z a^\dagger a^\dagger -  z^* a a)\}, \\
        z & = & r \exp(i \phi)
\eea Squeezed states are defined in terms of displacement and
squeeze operators as follows: \be
        \ket{{\alpha,z}}  = \Da \Sz \ket{0},
\ee where the ordering $\Da \Sz$ differs from $\Sz \Da$ by a
change of parameters, being however equivalent.

\subsection{Surprizes on $\pi^+$-$\pi^-$ Bose-Einstein correlations}
\label{ss:surprizes}

In 1991, Andrew, Pl\"umer and Weiner\cite{bbcapw} pointed out to
the surprising existence of a new quantum statistical correlation
among boson-antiboson pairs. The surprise was related to the fact
that this type of correlation involved particle-antiparticle
pairs, differently than the better known Bose-Einstein
Correlations (BEC), which occurred between two identical
particles. We can understand the origin of this effect in a simple
way in terms of creation and annihilation operators, taking the
$\pi^0$ case for illustration. For instance, the single-inclusive
distribution is written as
\begin{equation}
N_1({\mathbf k}_i) =  \omega_{{\mathbf k}_i}\frac{d^3N} {d{\mathbf
k}_i}
 =   \omega_{{\mathbf k}_i}  \langle \hat{a}^{\dagger}_{{\mathbf k}_i}
 \hat{a}_{{\mathbf k}_i} \rangle
\;, \label{spec1}\end{equation} and the two-particle distribution,
after the decomposition that follows from Wick's theorem, is
written as
\begin{eqnarray}
&& N_2({\mathbf k}_1,{\mathbf k}_2)  = \omega_{{\mathbf k}_1}
\omega_{{\mathbf k}_2} \langle \hat{a}^\dagger_{{\mathbf k}_1}
\hat{a}^\dagger_{{\mathbf k}_2} \hat{a}_{{\mathbf k}_2}
\hat{a}_{{\mathbf k}_1} \rangle
\nonumber\\
&=& \omega_{{\mathbf k}_1} \omega_{{\mathbf k}_1} \{ \langle
\hat{a}^{\dagger}_{{\mathbf k}_1}
\hat{a}_{{\mathbf k}_1} \rangle
\langle \hat{a}^{\dagger}_{{\mathbf k}_2} \hat{a}_{{\mathbf k}_2}
\rangle + \langle \hat{a}^{\dagger}_{{\mathbf k}_1}
\hat{a}_{{\mathbf k}_2} \rangle \langle
\hat{a}^{\dagger}_{{\mathbf k}_2} \hat{a}_{{\mathbf k}_1} \rangle
\nonumber\\
&& + \hskip.2cm \langle \hat{a}^{\dagger}_{{\mathbf k}_1}
\hat{a}^{\dagger}_{{\mathbf k}_2} \rangle \langle
\hat{a}_{{\mathbf k}_1} \hat{a}_{{\mathbf k}_2} \rangle \} \;.
\label{spec2}\end{eqnarray}

In Eq. (\ref{spec2}), the first term corresponds to the product of
the two single-inclusive distributions, the second one gives rise
to the Bose-Einstein identical particle correlation, reflecting
their last position just before being emitted. The third term,
absent in the $\pi^\pm \pi^\pm$ case, is the responsible for the
particle-antiparticle correlation (either $\pi^\pm \pi^\mp$ or
$\pi^0 \pi^0$, since $\pi^0$ is its own antiparticle). They are
related to the expectation value of the annihilation (creation)
operator, i.e., to
 $<\hat{a}^{(\dagger)}_{{\mathbf k}_1}
\hat{a}^{(\dagger)}_{{\mathbf k}_2}>\ne0$, analogous to what is
observed in two-particle squeezing in optics, where the averages
are estimated using a density matrix that contains squeezed
states, as briefly discussed in the previous section. Under the
conditions as those usually considered in femtoscopic analyses,
the last term in Eq. (\ref{spec2})
vanishes. However, it is non zero if the Hamiltonian of the system
is of the type $H=H_0+H_1$, where $H_0$ is the free part in the
vacuum, corresponding to final particles, and $H_1$ represents the
interaction of quasi-particles, resulting in an effective shift of
their masses. Alternatively, as in the pioneer work in Ref.
\cite{bbcapw}, this could be similar to having a {\sl chaotic
superposition of coherent states} and a density matrix containing
{\sl squeezed states}.

Although that initial discussion by Andreev, Pl\"umer and Weiner
was not entirely correct, they clearly pointed out that the
particle-antiparticle correlations would manifest themselves as an
enhancement above unity of the correlation function, i.e.,
$C(\pi^+\pi^-)>1$ and $C(\pi^0\pi^0)>1$, reflecting
particle-antiparticle quantum statistical effects.

In 1994, Sinyukov\cite{YS94}, also discussed a similar effect for
$\pi^+\pi^-$ and $\pi^0\pi^0$ pairs, claiming that it would be due
to inhomogeneities in the system, as opposed to homogeneity
regions in HBT, which comes from a hydrodynamical description of
the system evolution. He used Wick's theorem for expanding the
two-particle inclusive distribution in terms of bilinear forms.

In 1996, Andreev and Weiner\cite{andwei} elaborated further
their original idea. They considered that in high energy
collisions, a blob of strongly interacting pions, which was seen
as a liquid, was formed and later suffered a sudden breakup, so
that the pionic system having ground state and pioninc excitations
was rapidly converted into free pions. In the moment of
transition, they postulated that the in-medium creation and
annihilation operators ($\hat{b}^{\dagger},\hat{b}$) could be
related to the corresponding free ones
($\hat{a}^{\dagger},\hat{a}$), by means of a squeezing
transformation
\begin{eqnarray}
\hat{a} & = & \hat{b} \cosh(r) + \hat{b}^\dagger \sinh(r) \nonumber \\
\hat{a}^\dagger & = & \hat{b} \sinh(r) + \hat{b}^\dagger \cosh(r)
\; , \label{AWsqueez} \end{eqnarray} where
$r=\frac{1}{2}\ln(E_{fr}/E_{in})$ is a squeezing
parameter~\cite{Janszky1986}, $E_{fr}$ and $E_{in}$ are the
asymptotic (free) energy and the in-medium energy, respectively.


In the same year, Asakawa and Cs\"org\H{o}\cite{asacso} proposed a
similar structure to this previous approach, but relating
in-medium operators  to free ones by means of a two-mode Bogoliubov
transformation. 
They also
proposed to observe hadron mass modification in hot medium by
means of {\sl Back-to-Back Correlations},  
relating particle-antiparticle pair correlations
to two-mode squeezed states.

There were a few more tentative works by the two groups but the
correct approach was finally written in 1999, by Asakawa,
Cs\"org\H{o}, and Gyulassy\cite{acg}. The squeezing parameter they
proposed was somewhat similar to the one described by Eq.
(\ref{AWsqueez}).

\subsection{Back-to-back boson-antiboson correlations}
\label{ss:bBBC}

The formalism developed by Asakawa, Cs\"org\H{o} and Gyulassy for
squeezed bosons in an infinite medium can be summarized as
follows. The in-medium Hamiltonian, $H$, is written as
${H}  =   H_0 - \frac{1}{2} \int d {\mathbf x}~ d {\mathbf y}
\phi({\mathbf x}) \delta M^2({\mathbf x}-{\mathbf y})
\phi({\mathbf y}),
\;\;$
where $H_0  =  \frac{1}{2} \int d {\mathbf x} \left( \dot{\phi}^2+
|\nabla \phi|^2 + m^2 \phi^2  \right),
\;$
is the free Hamiltonian, in the rest frame of matter. The scalar
field, $\phi({\mathbf x})$, represents quasi-particles whose mass
is modified by the medium, and propagate in a momentum-dependent
way. The in-medium mass, $m_*$, is related to the vacuum mass,
$m$, via
$$ m_*^2({|{\mathbf k}|}) =  m^2 - \delta M^2({|{\mathbf k}|}).$$
The mass-shift is assumed to be limited to long wavelength
collective modes: $\delta M^2({|{\mathbf k}|}) \ll m^2$ if
$|{\mathbf k}| > \Lambda_s$. As a consequence, the dispersion
relation is modified to $\Omega_{\mathbf k}^2 =\omega^2_{\mathbf
k}-\delta M^2 (|{\mathbf k}|)$, where $\Omega_{\mathbf k}$ is the
frequency of the in-medium mode with momentum ${\mathbf k}$.

The in-medium, thermalized annihilation (creation) operator is
denoted by $b_{\mathbf k}$ ($b^\dagger_{\mathbf k}$), whereas the
corresponding asymptotic operator for the observed quantum with
four-momentum $k^{\mu}\, = \, (\omega_{\mathbf k},{\mathbf k})$,
$\omega_{\mathbf k}^2= m^2 + {\mathbf k}^2$ ($\omega_{\mathbf k}
> 0$) is denoted by $a_{\mathbf k}$ ($a^\dagger_{\mathbf k}$).
These operators are related by the Bogoliubov transformation,
i.e., $ a_{{\mathbf k}_1} = c_{{\mathbf k}_1} b_{{\mathbf k}_1} +
s^*_{-{\mathbf k}_1} b^\dagger_{-{\mathbf k}_1} $, which is
equivalent to a squeezing operation. For this reason, $r_{\mathbf
k}$ is called mode-dependent squeezing parameter. The relative and
the average pair momentum coordinates are written as $q^0_{1,2}=
\omega_1-\omega_2,$ ${\mathbf q}_{1,2}={\mathbf k}_1- {\mathbf
k}_2$, $E_{i,j}=\frac{1}{2}(\omega_i+\omega_j)$, and ${\mathbf
K}_{1,2}=\frac{1}{2}({\mathbf k}_1+ {\mathbf k}_2)$. For
shortening the notation, the squeezed functions are denoted by
$c_{i,j} = \cosh[r({i,j},x)]$ and $s_{i,j} = \sinh[r({i,j},x)]$,
where
\begin{eqnarray}
r(i,j,x) &=& \frac{1}{2}\log \left[( K^{\mu}_{i,j} u_\mu (x))/
(K^{* \nu}_{i,j}(x) u_\nu(x) ) \right]
\nonumber\\
&=& \frac{1}{2}
\left[\frac{\omega_{k_i}(x)+\omega_{k_j}(x)}{\Omega_{k_i}(x)+
\Omega_{k_j}(x)} \right]
\end{eqnarray}
is the squeezing parameter. Also, $n_{i,j}$ is the density
distribution, which is taken as the Boltzmann limit of the
Bose-Einstein distribution, i.e., $ n^{(*)}_{i,j} (x) \approx
\exp{\{ - [K^{(*)^\mu}_{i,j} u_\mu(x) - \mu (x)]/T(x)\}} $, where
the symbol ($^*$) implies the use of in-medium mass, whereas it is
absent if there is no mass-shift.

In the cases of $\pi^0 \pi^0 $ or $\phi \phi$ correlations, where
the boson is its own anti-particle, the full correlation function
consists of a HBT part (related to the chaotic amplitude,
$G_c(1,2)$) together with a BBC portion (related to the squeezed
amplitude, $G_s(1,2)$), as shown below
\begin{eqnarray}
&&C_2({\mathbf k}_1,{\mathbf k}_2)  =
 \frac{N_2({\mathbf k}_1,{\mathbf k}_2)}
 {N_1({\mathbf k}_1) N_1({\mathbf k}_2)}
\nonumber \\
&&= 1 + \frac{| G_c(1,2) |^2}{G_c(1,1) G_c(2,2) } + \frac{|
G_s(1,2) |^2}{G_c(1,1) G_c(2,2) }. \label{fullcorr}
\end{eqnarray}
The invariant single-particle and two-particle momentum
distributions given by
\begin{eqnarray}
G_c(i,i) &=& G_c(k_i,k_i) = N_1({\mathbf k}_i)= \omega_{{\mathbf
k}_i}  \langle \hat{a}^{\dagger}_{{\mathbf k}_i} \hat{a}_{{\mathbf k}_i} \rangle, \nonumber\\
G_c(1,2)&=& \sqrt{\omega_{{\mathbf k}_1} \omega_{{\mathbf k}_2} }
\langle \hat{a}^\dagger_{{\mathbf k}_1} \hat{a}_{{\mathbf
k}_2}\rangle,
\nonumber \\
G_s(1,2) &=& \sqrt{\omega_{{\mathbf k}_1} \omega_{{\mathbf k}_2}
}\langle \hat{a}_{ {\mathbf k}_1} \hat{a}_{{\mathbf k}_2} \rangle
.\label{rand}\end{eqnarray}

\bigskip
For an infinite, homogeneous and thermalized medium, the part of
the full correlation function in Eq. (\ref{fullcorr}),
corresponding to the Back-to-Back Correlation is written as
\begin{equation}
 C_{bosons}(\mk, -\mk)  =  1 + \frac{|c_{\mk} s^*_{\mk} n_{\mk} +
c_{-\mk} s^*_{-\mk} \left(n_{-\mk} + 1\right)|^{\, 2} } {n_1(\mk)
\, n_1(-\mk)}. \label{e:c2b} \end{equation}

The effects of finite size on BBC are considered afterwards in the
text. In Ref.\cite{acg}, the influence of finite emission times is
discussed, observing that the BBC in this case is suppressed when
compared to the instant emission.
Nevertheless, it was also shown that the maximum of the BBC for
each value of $|\vec{k}|$, corresponding to $C_2({\mathbf
k},-{\mathbf k})$, still attained significant magnitude, in spite
of considering the time suppression. We will see this more
explicitly later.

\subsection{Back-to-back correlations for fermions}
\label{ss:fBBC}

Previously, the BBC was shown as a different type of correlation
between boson-antiboson pairs, occurring if their masses were
shifted. In 2001, T. Cs\"org\H{o}, Y. Hama, G. Krein, P. K. Panda
and S. S. Padula demonstrated that a similar correlation existed
between fermion-antifermion pairs\cite{pchkp}, if their masses
were modified in a thermalized medium. In the femtoscopic type of
correlations, identical bosons have an opposite behavior as
compared to identical fermions, resulting from the fact that
quantum statistics suppresses the probability of observing pairs
of identical fermions with nearby momenta, while it enhances such
a probability in the case of bosons. However, regarding the
Back-to-Back Correlations resulting from squeezed states, a very
different situation occurs: fermionic BBC are positive and similar
in strength to bosonic BBC. Besides, contrary to the the
femtoscopic correlations, the BBC are unlimited.

The expressions in the fermion BBC case are similar to
Eq.(\ref{spec1}) and (\ref{spec2}), \vskip-.5cm\begin{equation}
N_1({\mathbf k}_i) =  \omega_{{\mathbf k}_i}  \langle
a^{\dagger}_{{\mathbf k}_i}
 a_{{\mathbf k}_i} \rangle \; \; ; \; \;
 \tilde{N}_1({\mathbf k}_i) = \omega_{{\mathbf k}_i}
 \langle \tilde{a}^{\dagger}_{{\mathbf k}_i}
 \tilde{a}_{{\mathbf k}_i} \rangle \;, \label{spec1f}\end{equation}
\vskip-.8cm\begin{equation} N_2({\mathbf k}_1,{\mathbf k}_2)  =
\omega_{{\mathbf k}_1} \omega_{{\mathbf k}_2} \langle
a^\dagger_{{\mathbf k}_1} \tilde{a}^\dagger_{{\mathbf k}_2}
\tilde{a}_{{\mathbf k}_2} a_{{\mathbf k}_1} \rangle \; .
\label{spec2f}\end{equation} \vskip-.2cm In the above expressions,
$\langle\hat{O}\rangle$ denotes the expectation value of the
operator $\hat{O}$ in the thermalized medium and $a^\dagger, a,
\tilde{a}^\dagger, \tilde{a}$ are, respectively, creation and
annihilation operators of the free baryons and antibaryons of mass
$M$ and $\omega_k=\sqrt{M^2+|\vec k^2|}$, which are defined
through the expansion of the baryon field operator as $\Psi(\vec
x) = (1/V)\sum_{\lambda,\lambda',\vec k} (u_{\lambda,\vec k}
a_{\lambda,\vec k}+v_{\lambda',-\vec k} a^\dagger_{\lambda',-\vec
k}) e^{i \vec k.\vec x}$; $V$ is the volume of the system,
$u_{\lambda,\vec k}$ and $v_{\lambda',-\vec k}$ are the Dirac
spinors, where the spin projections are
$\lambda,\lambda'=1/2,-1/2$. The in-medium creation and
annihilation operators are denoted by $b^\dagger, b,
\tilde{b}^\dagger, \tilde{b}$. While the $a$-quanta are observed
as asymptotic states, the $b$-quanta are the ones thermalized in
the medium. They are related by a fermionic Bogoliubov-Valatin
transformation,
\begin{equation}
\left(
\begin{array}{c}
a_{\lambda,{\mathbf k}} \\
\tilde{a}^{\dagger}_{\lambda',-{\mathbf k}} \end{array} \right) =
\left(
\begin{array}{cc}
c_{\mathbf k} &
\frac{f_{\mathbf k}}{|f_{\mathbf k}|} \,s_{\mathbf k} \,A  \\
-\frac{f^*_{\mathbf k}}{|f_{\mathbf k}|}\, s^*_{\mathbf k}\,
A^{\dagger} & c^*_{\mathbf k}
\end{array}
\right) \left(
\begin{array}{c}
b_{\lambda,{\mathbf k}} \\
\tilde{b}^{\dagger}_{\lambda',-{\mathbf k}}
\end{array}      \right ),
\label{BVtransf}
\end{equation}
here $c_1=\cos r_1$, $s_1=\sin r_1 $, and
\begin{equation}
\tan(2 r_1) = - \frac{|{\bf k}_1| \Delta M({\bf k}_1) }
{\omega({\bf k}_1)^2 - M \Delta M({\bf k}_1) } \label{fk}
\end{equation}
is the fermionic squeezing parameter. Note that in the fermionic
case, the squeezing parameter is the coefficient of sine and
cosine functions, differently than the bosonic cases in which
appeared their hyperbolic counterparts. In Eq.~(\ref{BVtransf})
$A$ is a $2~\times~2$ matrix with elements
$A_{\lambda_1\lambda_2}=
\chi^{\dagger}_{\lambda_1}\sigma\cdot{\hat{\bf k}}_1\tilde
\chi^{\phantom\dagger}_{\lambda_2}$, where $\hat{\bf k}_1 = {\bf
k}_1/|{\bf k}_1|$,  $\chi$ is a Pauli spinor and $\tilde \chi=
-i\sigma^2 \chi$. Since $r$ is real in the present case, we drop
the complex-conjugate notation in what follows.

\begin{figure}[!hbt25]
\includegraphics*[angle=0,width=5.5cm]{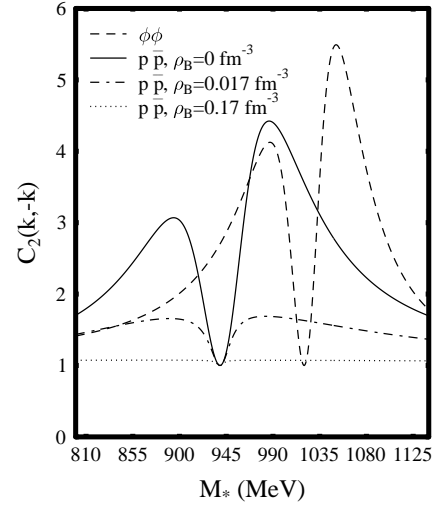}
\caption{The plot shows back-to-back correlations of $\bar{p}p$
(fBBC) and of $\phi$-meson pairs (bBBC), as a function of the
in-medium modified ($p$ or $\phi$) mass, $m_*$, for $|\vec k|=800$ MeV/$c$.
The dependence of the fBBC on the net baryon density is shown for
three values of $\rho_B$. In both cases $T=140$ MeV and $\Delta
t=2$ fm/$c$. The plots were extracted from Ref.\cite{pchkp}. }
\label{bbc1}\end{figure}

In order to evaluate the thermal averages above, the system is
modelled as a globally thermalized gas o quasi-particles
(quasi-baryons). In this description, the medium effects are taken
into account through a self-energy function, which, for a
spin-$\frac{1}{2}$ particle (we will focus on proton and
anti-proton pairs), under the influence of mean fields in a
many-body system, can be written as $\Sigma = \Sigma^s + \gamma^0
\Sigma^0 + \gamma^i \Sigma^i$. In this expression, $\Sigma^0$ is a
weakly momentum dependent function which, for locally thermalized
systems that we are considering, has the role of shifting the
chemical potential, i.e., $\mu_* = \mu - \Sigma^0$. The vector
part is very small and is neglected. The scalar part can be
written as $\Sigma^s = \Delta M({\bf k})$. Within these
approximations the system can be described with a
momentum-dependent in-medium mass, $m_*({\bf k}) = m - \Delta
M(|{\bf k}|)$.

We are mainly interested here in the study of the squeezed
correlation function, which corresponds to considering only the
joint contribution of the first and third terms of the rhs of Eq.
(\ref{fullcorr}). In the fermionic case and for an infinite,
homogeneous thermalized medium, the BBC part of the correlation
function is written as
\begin{eqnarray}
&&\!\!\!\!\!\!\!\!\!\!\!\!C_{fermions}^{(+-)}({\bf k}_1,-{\bf
k}_1) = 1 + \; [1+(2\Delta t \; \omega_{\bf k})^2]^{-1} \; \times
\nonumber\\
&&\!\!\!\!\!\!\!\!\!\!\!\!\{\frac{(1 - n_{\bf k} - \tilde{n}_{\bf
k})^2 (c_{\bf k} s_{\bf k})^2} { \left[ c_{\bf k}^2 n_{\bf k} +
s_{\bf k}^2 (1-\tilde{n}_{\bf k})
    \right]
 \left[ c_{\bf k}^2 \tilde{n}_{\bf k} + s_{\bf k}^2  (1-n_{\bf k})
    \right] }\},\; \label{BBCFa}
\end{eqnarray}
where $n_{\bf k} =  \frac{1}{\exp \left[(\Omega_{\bf k} -
\mu_*)/T\right] +1} \; ; \; \tilde{n}_{\bf k}  =  \frac{1}{\exp
\left[(\Omega_{\bf k} + \mu_*)/T\right] +1}$ in terms of which the
net baryonic density is written as $\rho_B = (g/V)\sum_{\bf
k}~\bigl( n_{\bf k} - \tilde{n}_{{\bf k}}\bigr)$. In Eq.
(\ref{BBCFa}) we have included a more gradual freeze-out by means
of a finite emission interval, similarly to what was done in
Ref\cite{acg}, which has the effect of suppressing the BBC signal.

For a numerical study of the fermionic back-to-back correlations,
fBBC, we considered, for simplicity, momentum independent
in-medium masses, i.e., $m_* = M - \Delta M$.
In Fig. (\ref{bbc1}) we show fBBC for $\bar{p}p$ pairs as a
function of the in-medium mass $m_*$, for three values of the net
baryonic density $\rho_B$: for the normal nuclear matter, one
tenth of this value and for the baryon free region, i.e.,
$\rho_B=0$. We show in the same plot results for the bosonic case,
bBBC, corresponding to $\phi$ meson pair, whose mass is close to
the proton mass and was the example used in Ref.\cite{acg}.

We see from Fig.(\ref{bbc1}) that fBBC and bBBC are, indeed, both
positive correlations, with similar shape, and of the same order
of magnitude. We also observe that fBBC is strongly enhanced for
decreasing net baryonic density, being maximal for $\rho_B \approx
0$, i.e., for approximately equal baryon and anti-baryon
densities.


\subsection{Flow effects on back-to-back correlations}
\label{ss:flow-BBC}

In the previous discussions, an infinite and homogeneous medium
was considered. However, we know that the systems produced in high
energy collisions, including the ones at RHIC, have finite sizes.
Thus, it would be important to test if the BBC signal would
survive when more realistic spatial and dynamical hypotheses were
considered. For pursuing this purpose, we studied the effects on
the squeezing parameter and on the back-to-back correlation of a
finite size medium moving with collective velocity\cite{pkchp}.
For this, a hydrodynamical ensemble was adopted, in which the
amplitudes $G_c$ and $G_s$ in Eq. (\ref{fullcorr}) and
(\ref{rand}) were extended to the special form derived by Makhlin
and Sinyukov \cite{YS94},
\begin{eqnarray}
G_c(1,2)&=&\frac{1}{(2 \pi)^3 } \int d^4\sigma_{\mu}(x)
K_{1,2}^{\mu} e^{i q_{1,2} \cdot x} \{|c_{1,2}|^2 n_{1,2}
\nonumber\\
&& \hskip1cm + \;\;\; | s_{-1,-2}|^2  (n_{-1,-2} + 1) \},
\label{e:gc}
\end{eqnarray}
\begin{eqnarray}
G_s(1,2)\!\!&=&\!\!\frac{1}{(2 \pi)^3 } \int d^4\sigma_{\mu}(x)
K_{1,2}^{\mu} e^{2 i  K_{1,2} \cdot x} \{ s^*_{-1,2} c_{2,-1}
n_{-1,2}
\nonumber\\
&& \hskip1cm + \;\;\; c_{1,-2} s^*_{-2,1} (n_{1,-2} + 1) \}.
\label{e:gd}
\end{eqnarray}

\smallskip\smallskip
\noindent In Eq.(\ref{e:gc}) and (\ref{e:gd}) $d^4\sigma^{\mu}(x)
= d^3\Sigma^{\mu}(x;\tau_f)\, F(\tau_f) d\tau_f $ is the product
of the normal-oriented volume element depending parametrically on
$\tau_f$ (the freeze-out hyper-surface parameter) and the
invariant distribution of that parameter $F(\tau_f)$. We consider
two possibilities: i) an instant freeze-out, corresponding to
$F(\tau)=\delta(\tau-\tau_0)$; ii) an extended freeze-out, with a
finite emission interval, with
$F(\tau)=[\theta(\tau-\tau_0)/\Delta t] e^{-(\tau-\tau_0)/\Delta
t}$. These cases lead, after performing the integration in $d
\tau$ in Eq. (\ref{e:gc}) and (\ref{e:gd}) with weight ($E_{i,j}
\;e^{-i 2 E_{i,j} \tau}$), respectively to: i)
$(\omega_i+\omega_j) \; e^{- i (\omega_i+\omega_j) \tau_0}$; ii)
$(\omega_i+\omega_j)[1+[(\omega_i+\omega_j)^2\Delta t^2]^{-1/2}$.

\begin{figure}[!hbt25]
\resizebox{17pc}{!}{\includegraphics{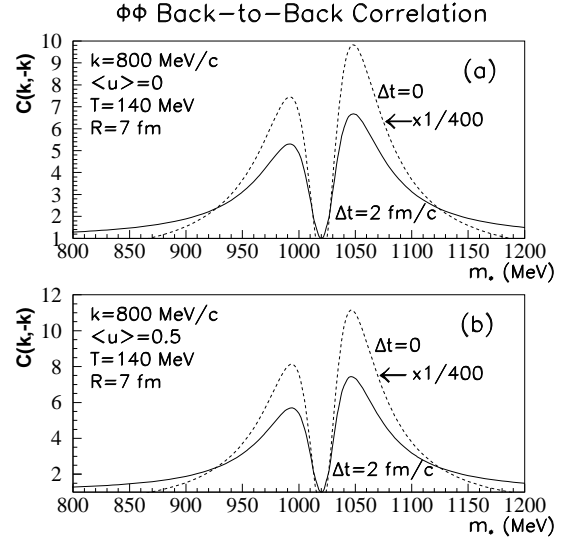}}
\caption{The effect of a finite emission interval on the back to
back correlation function, as compared to instant emission, is
illustrated by the two plots. The dashed curves have been reduced
by a factor of $400$, and the solid curves correspond to the
suppression by a finite emission duration, of about $\Delta t
\simeq 2$~fm/$c$. The plot in (a) shows this effect in the absence
of flow. The plot  in (b) shows the corresponding result when flow
is included, with $<\!\!u\!\!>\!=0.5$; the other parameters
adopted to produce the curves are $R=7$ fm/$c$, $T=140$~MeV. 
The plots were extracted from Ref.\cite{pkchp}. }
\label{bbc2}\end{figure}


We estimate the geometrical and dynamical effects for moderate
flow on the BBC in the bosonic case, considering the in-medium
changes of $\phi$-mesons as illustration.

For small mass shifts, i.e. $\frac{(m-m_*)}{m} \ll 1$, the flow
effects on the squeezing parameter are of fourth order, i.e.,
${\cal O} \left( \frac{Kin. \; energy}{m} \right) (\frac{\delta
m^2}{m^2})$. As a consequence, the flow effects on $r_{i,j}$ can
be neglected, and the factor $c_{i,j}$ and $s_{i,j}$ become flow
independent, although they could still depend on the coordinate
$r$ through the shifted mass, $m_*$ (e.g., through $T(x)$, as in
hydrodynamics), which is not considered here.

\begin{figure}[!hbt26]
\resizebox{18pc}{!}{\includegraphics*{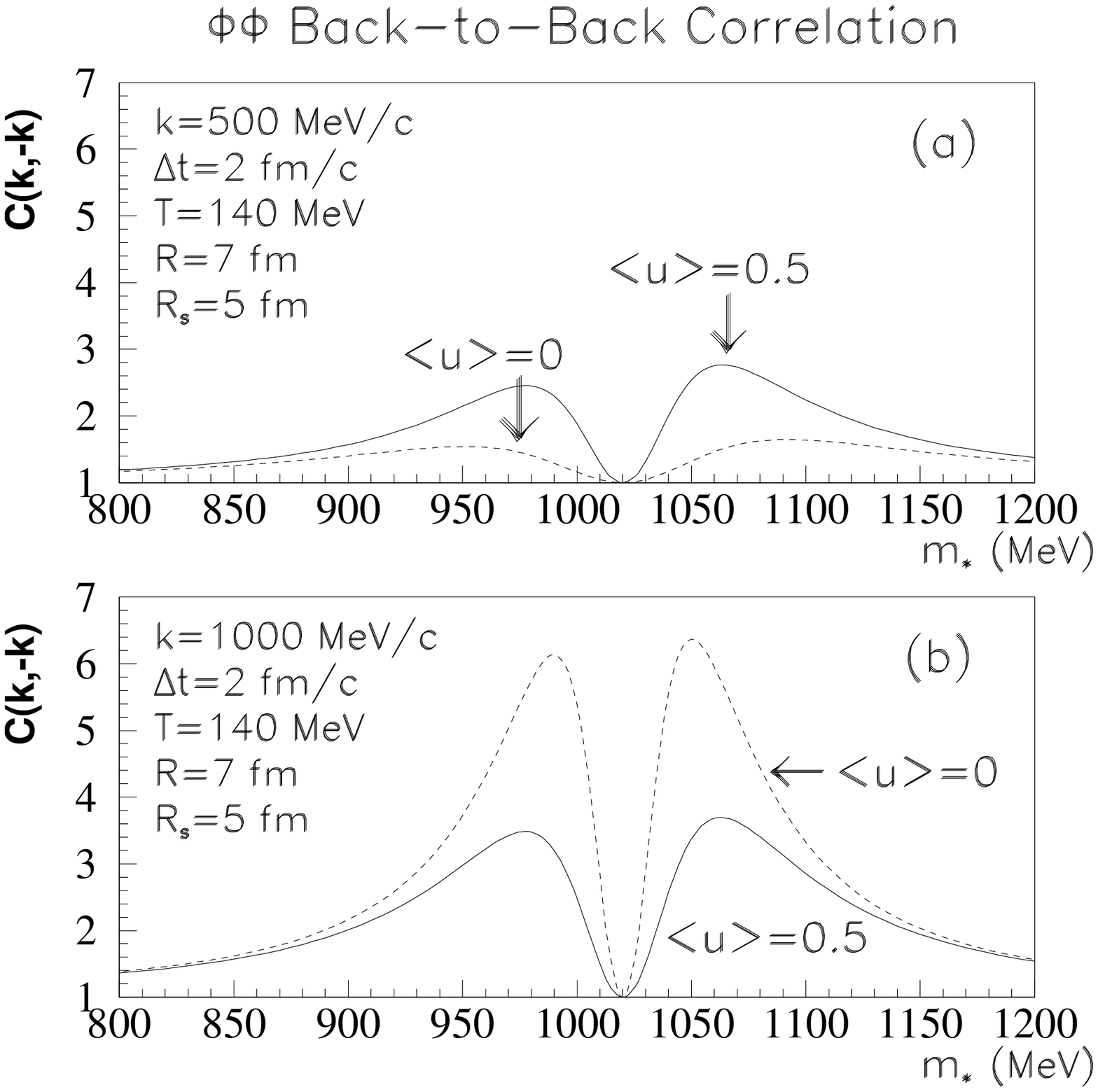}} 
\resizebox{18pc}{!}{\includegraphics{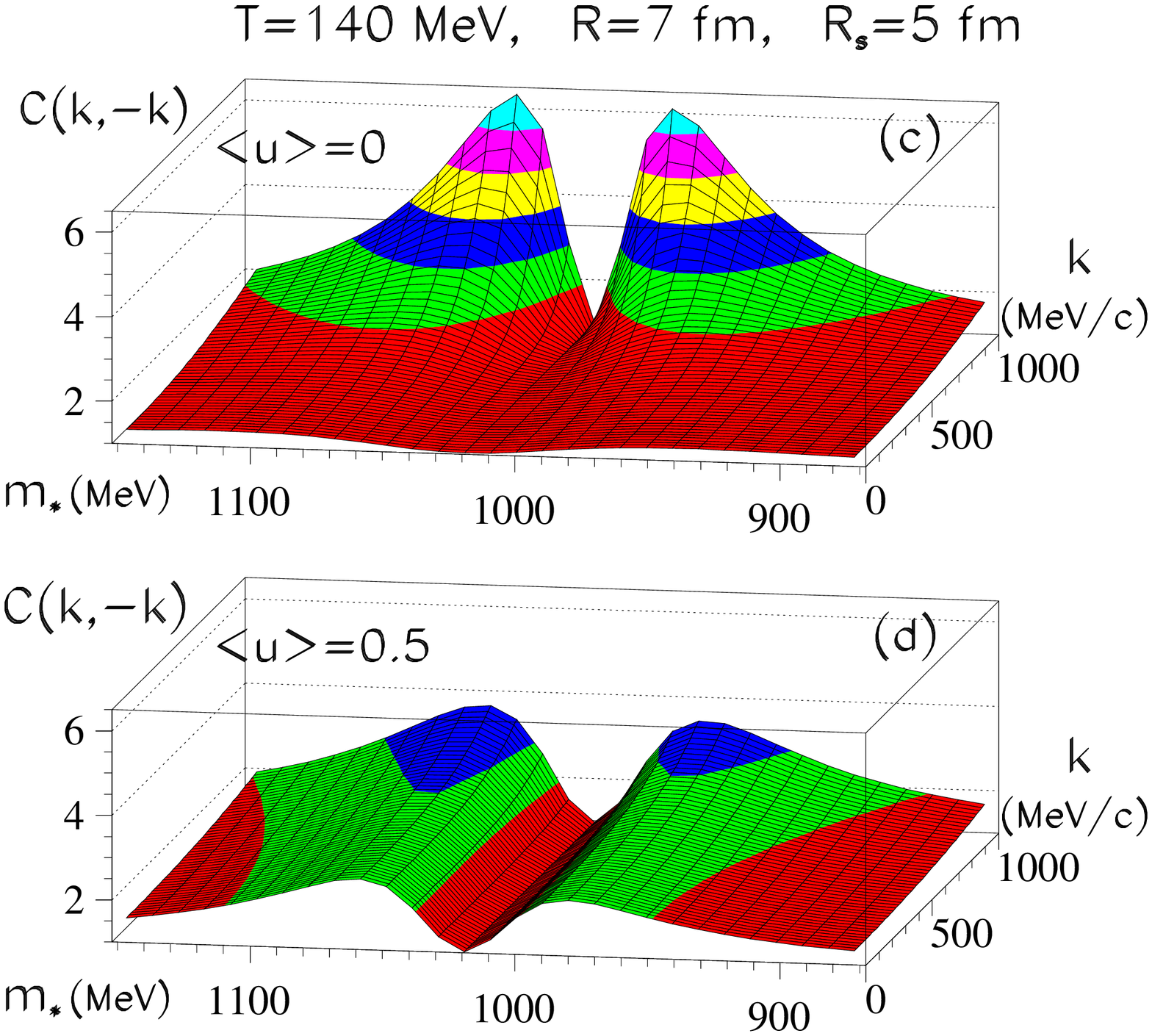}} \caption{The
plots in this panel are similar to the ones on Fig.~\ref{bbc2}.
The main difference is that here we assumed that the mass-shift
occurred only in a smaller part of the system volume. The
back-to-back correlation is shown as a function of the shifted
mass $m_*$ on top, and as function of both $m_*$ and the momentum
of each particle ($\mk_1=-\mk_2=\mk$), on bottom. 
The plots in parts (a) and (b) illustrate better the behavior of
the BBC signal seen in parts (c) and (d), for $|\mk| =
500$~MeV/$c$ and for $|\mk| = 1000$~MeV/$c$, respectively. In both
cases, the dashed curve corresponds to $\langle u \rangle = 0$ and
the solid curve, to $\langle u \rangle =0.5$. In (c), the 3-D plot
without flow ($\langle u \rangle =0$) was considered, whereas in
(d) a radial flow with $v = \langle u \rangle r/R= 0.5$ was
included. The plots were extracted from Ref.\cite{pkchp}.}
\label{bbc3}\end{figure}

For the sake of simplicity, and trying to keep the results as
analytic as possible (for details, see Ref. \cite{pkchp}), we made
the hypothesis that the mass-shift was independent on the position
within the fireball. We further assumed that this last one had a
sharp boundary, i.e., $\delta m=0$ on the surface, and also the
density vanishes outside the system volume. The spatial
integration in Eq. (\ref{e:gc}) and (\ref{e:gd}) extends over the
region where the mass-shift is non-vanishing,  which is {\it not}
infinitely large. For instance,  in relativistic heavy ion
collisions is a finite region $V \approx R^3 \approx (5-10)^3$
fm$^3$. We should keep in mind that the vacuum term in the
integrand vanishes outside the mass-shift region, since it is
proportional to $s_{i,j}$, which is identically zero in that
region. On the other hand, the terms proportional to
$n^{(*)}_{i,j} (x)$ are finite.
Being so, we can extend the integration in Eq. (\ref{e:gc}) and
(\ref{e:gd}) to infinity and, without much loss of generality, we
can choose for $V$ a Gaussian profile, $\exp[-{\mathbf
r}^2/(2R^2)]$. This study was performed considering two
situations: in one of them, the mass-shift occurs in the entire
system region, which is considered as 3-D Gaussian with a circular
cross-sectional area of radius $R=7$ MeV/$c$  at its width. The
second case considers this volume split in two regions, the
mass-shift occurring only in the internal one, with $R=5$ MeV/$c$.
Although we adopted several simplifying hypotheses for reducing
the problem complexity and treat it analytically, the resulting
expressions are still intricate, so that a graphical presentation
of the results is more beneficial to the reader. A complete
discussion with full analytical and numerical results can be found
in Ref.\cite{pkchp}. 

Therefore, Fig. (\ref{bbc2}) and Fig. (\ref{bbc3}) summarize the
main results of that study. The panel in Fig. (\ref{bbc2}) shows
the case where the mass shift extends over the entire system
region. It clearly illustrates the dramatic suppression effect of
a finite emission duration, since the curves corresponding to
instant emission had to be reduced by a factor $1/400$ in order to
fit in the same plots . We can also see that the effect of flow,
within the approximations used, also reduces the signal but in a
moderate way.

For illustrating other interesting features found in this
preliminary analysis and shown in Fig.(\ref{bbc3}), we chose the
case with two-regions, as briefly delineated above. It also allows
for some comparison with the result on Fig.(\ref{bbc2}). In parts
(a) and (b) of Fig.(\ref{bbc3}), we can see that, depending on the
momenta of the back-to-back pairs suffering the squeezing
correlation, the signal is weaker when flow is present (for $|\mk|
\simeq 1000$ MeV/$c$), almost unaffected by flow  (for $|\mk|
\simeq 750$ MeV/$c$), or even slightly stronger in the presence of
moderate flow (for $|\mk| < 500$ MeV/$c$), for the set of
parameters chosen in these calculations. If flow is absent,
however, a monotonically increase can be observed with increasing
values of the pair momenta, $|\mk|$. A more complete view of the
behavior of the maximum of the $\phi\phi$ back-to-back correlation
can be seen in the parts (c) and (d) of Fig.(\ref{bbc3}). It is
evident that BBC correlation function has a steeper growth with
momentum in the no flow case, for the same values of the shifted
mass, $m_*$. The moderate flow picture considered in this study
still causes the growth of the BBC with $|\mk|$, but in a
considerably smaller rate. With this broader panorama in mind, it
is easier to understand the plots (a) and (b) of Figs.
(\ref{bbc2}) and (\ref{bbc3}). Finally, we see that the strength
of the signal is directly proportional to the size where the
mass-shift occurs. Although the values of $|\mk|$ were not the
same in those two figures, but already knowing that the signal
grows with increasing $|\mk|$, we see that the strength in Fig.
(\ref{bbc2}), where the system size is $R=7$ fm, is bigger than
the corresponding ones in Fig. (\ref{bbc3}), in which the
squeezing region has $R_s=5$ fm, even for $|\mk|=1000$ GeV/$c$.

Naturally, the above study was mainly a first step towards better
understanding the nature and conditions of survival of the
squeezing correlation, since it involved several approximations in
its derivation. In particular, it was supposed that the squeezing
occurred homogeneously throughout the system region, and
independently on the particle momentum. Next steps will require a
modelling instead of this assumption, which will provide the
volume dependence of the squeezing in a more realistic way.
Besides, we have presented merely the behavior of the maximum of
the Back-to-Back Correlation function. Is is under investigation
how the signal behaves for finite systems and small values of the
average pair momenta, which hopefully will furnish the optimized
form of the BBC signal that should be looked for experimentally.

\subsection{Chiral dynamics and back-to-back correlations}
\label{ss:chiral-BBC} This sub-section is based on
ref.~\cite{Randrup:2000je}. The environment generated in the
mid-rapidity region of a high-energy nuclear collision might endow
the pionic degrees of freedom with a time-dependent effective
mass. This implies two-mode squeezing, but with a time-dependent
squeezing parameter. Its specific evolution provides a mechanism
for the production of back-to-back charge-conjugate pairs of soft
pions which may present an observable signal of the 
non-equilibrium dynamics of the chiral order parameter. The
suddenness of the transition to the asymptotic quanta is a
condition that is released in this approach. The important point
of the numerical investigations along such a model assumptions
were that suddenness is not a mandatory requirement.
 The BBC signal seems to be strong enough to survive even
the time-dependence of the effective mass, caused by chiral
dynamics, resulting in an adiabatic mass variation, modelled by
mass oscillations within an exponentially decreasing envelope.

\subsection{Experimental constrains on chiral dynamics as a suggested explanation of the
RHIC HBT puzzle}
Recently, Cramer, Miller and collaborators published an interesting work, that included
a relativistic quantum mechanical treatement of opacity and refractive effects,
that allowed for a reproduction of STAR two-pion (HBT) correlation data and pion spectra
in Au+Au collisions at $\sqrt{s_{NN}} = 200$ GeV colliding energies at RHIC.
This investigation suggested, that an attractive, real part of the optical potential is
the critical element needed to reproduce the transverse mass dependence of the sidewards and the
outwards HBT radii at RHIC. This optical potential represents the strength of the interactions
between the pions and the medium. Chiral dynamics in the model leads to 
a temperature and density dependent in-medium modification of the pion (pole as well as screening) mass.
The results were found to be consistent
with a system, that had a restored chiral symmetry~\cite{Cramer:2004ih}.

Cramer, Miller and collaborators have suggested to check experimentally 
various new phenomena to see if their explanation is correct: they have argued
that a pionic version of the Ramsauer resonances lead to peaks in the HBT radius parameters
$R_o$ and $R_s$ in the low momentum region, 15 MeV $< p_t < 65 $ MeV,
and have also predicted a peaking behaviour in the same region in the transverse momentum  spectrum of pions
\cite{Cramer:2004ih,Miller:2005ji,Cramer:2006em}.

Apparently, there are various other possible tests that can be performed to see if this
connection of pion HBT radii to chiral dynamics is a unique, and fully correct explanation, or not.
For example, if in medium mass modifications of pions is the reason for the effect,
which depends on the interactions of pions with the medium, then similar interactions
but of different strength should be present between kaons and the medium, hence 
kaon and pion HBT radii are not expected to show a similar transverse mass dependence. On the other hand,
if the HBT radii are equal due to asymptotic properties of exact solutions of relativistic
hydrodynamics, as proposed by the Buda-Lund hydro model~\cite{BL-indication}, 
kaons, pions, and all other heavier particles should show the same effective, 
and transverse mass dependent, approximately
spherically symmetric source size at large transverse masses,
($R_{out} \simeq R_{side} \simeq R_{long} \propto 1/\sqrt{m_t}$).
So the calculation of the transverse momentum dependence of the pion HBT radii above the
presently limited to 600 MeV upper limit in transverse momentum from this model,
and the comparison of this value to the PHENIX results on these radii up to 1.2 GeV
in transverse momentum, and a theoretical and experimental combined effort to do a similar
study for kaons could help to clarify the case.

However, such an investigation inevitably needs a cooperation between theorists and experimentalists,
and any possible conclusions will have  a definite model dependence. 
From the experimental point of view, it is better to find a clear-cut effect, which has to exist,
if indeed chiral dynamics is the key element in the success of the Cramer-Miller approach.
As should be clear from our presented review on squeezed correlations,
back-to-back correlations of particle-antiparticle pairs were shown to exist in all scenarios
that discuss in-medium mass modification of hadrons, regardless of the details of the calculation
(bosons or fermions, expansion, finite duration of particle emission, in-medium mass modification
only in a fraction of the total volume, time dependent in-medium mass modification due to chiral dynamics).
Thus the experimental determination of the in-medium mass modification with the help of 
back-to-back particle anti-particle correlations could give  a clear-cut answer, if the
success of the Cramer-Miller model is indeed due to a chiral phase transition, or perhaps due to
other, less unique elements of their  model.

\section{SUMMARY: A NEW CORRELATION SIGNATURE OF A  FREEZE-OUT FROM A SUPERCOOLED
QUARK-GLUON PLASMA}
\label{s:noneq-signature}

Currently available 
correlation data
seem to exclude the possibility of a strong first order phase
transition in Au+Au collisions at RHIC and in Pb+Pb collisions at
top CERN SPS energies. The proposed signature of a possible second
order QCD phase transition has not yet  been investigated  in
detail, the excitation function of the L\'evy index of stability
needs to be determined experimentally. From lattice QCD
calculations there is an indication that, in case such a critical
point is reached at a certain value of the center of mass energy
then, above this point, a cross-over type of transition should be
expected. Presently, a cross-over type of equilibrium
hadronization, or, a sudden non-equilibrium hadronization
mechanism are both possible scenarios for hadron production in
ultra-relativistic heavy ion collisions at RHIC. We have focused
here on the latter possibility, and discussed a quantum optical
characteristics of such a scenario, arguing that it leads to
vanishing in-medium mass modifications of hadrons. As a
consequence, the back-to-back correlated particle-antiparticle
pairs should disappear at the onset of such a sudden deconfinement
transition, characteristic of quark-gloun plasma hadronization and
freeze-out from a deeply supercooled, negative pressure state.
These back-to-back correlated particle-antiparticle pairs
should on the other hand appear, if a chiral phase transition 
is present in $\sqrt{s_{NN}}= 200$ GeV Au +Au collisions
at RHIC. Thus they can also be utilized as model independent experimental
tools to test the validity of the Cramer-Miller approach.

\bigskip
\noindent\textbf{Acknowledgments:} T. Cs. would like to express
his gratitude to professors S. S. Padula, Y. Hama and M. Hussein
for the kind invitation and hospitality in S\~ao Paulo, and for
their  organizing a series of  outstanding conferences in Brazil.
Both authors gratefully acknowledge the support from FAPESP
covering T. Cs\"org\H{o}'s visit (Proc. N. 03709-7 and {\sl
Projeto Tem\'atico} N. 2004/10619-9), as well as FAPESP (Proc. N.
03313-6) and CAPES (Proc. PAEX N. 0290/06-8) for partially
supporting the II Workshop on Particle Correlations and Femtoscopy
(WPCF 2006).


\end{document}